\documentclass{article}
\usepackage{spconf,amsmath,graphicx,url,times}
\usepackage{color}
\usepackage{enumitem}
\usepackage{balance}
\usepackage{multirow}

\usepackage[]{algorithm2e}
\usepackage{wasysym}
\usepackage{bm}
\usepackage{url,cite}
\usepackage{enumitem} % tune margin for \begin{itemize}
\usepackage{pifont}% http://ctan.org/pkg/pifont
\usepackage{multicol}
\title{Bootstrapping deep music separation from \\ primitive auditory grouping principles}

\name{Prem Seetharaman$^{1}$, Gordon Wichern$^2$, Jonathan Le Roux$^2$, Bryan Pardo$^1$\thanks{This work has made use of the Mystic (Programmable Systems Research Testbed to Explore a Stack-WIde Adaptive System fabriC) NSF-funded infrastructure at Illinois Institute of Technology, NSF award CRI-1730689.}}
\address{
$^1$Northwestern University, Evanston, IL, USA\\
$^2$Mitsubishi Electric Research Laboratories (MERL), Cambridge, MA, USA\\
}

\begin{document}

\ninept
\maketitle

\begin{sloppy}

\begin{abstract}
Separating an audio scene such as a cocktail party into constituent, meaningful components is a core task in computer audition. Deep networks are the state-of-the-art approach. They are trained on synthetic mixtures of audio made from isolated sound source recordings so that ground truth for the separation is known. However, the vast majority of available audio is not isolated. The brain uses primitive cues that are independent of the characteristics of any particular sound source to perform an initial segmentation of the audio scene. We present a method for bootstrapping a deep model for music source separation without ground truth by using multiple primitive cues. We apply our method to train a network on a large set of unlabeled music recordings from YouTube to separate vocals from accompaniment without the need for ground truth isolated sources or artificial training mixtures.
\end{abstract}

\begin{keywords}
auditory scene analysis, deep learning, cocktail party problem, self-supervised learning, bootstrapping
\end{keywords}

\section{Introduction}
\label{sec:intro}
A fundamental problem in computer audition is audio source separation, the act of isolating sound producing sources (or groups of sources) in an audio scene. Source separation is important for building machines that can perform perceptual audio tasks on realistic audio input (e.g., mixtures of sounds) with human-level performance. Deep models are the current state-of-the-art for source separation of mixtures of speech and music \cite{stoter2018sisec}. These models are trained on thousands of synthetic mixtures of isolated recordings of musical instruments and voices. Using synthetic mixtures guarantees that the ground truth isolated sources are known. However, most sounds in the world do not occur in recording studios and most available audio recordings (e.g., field recordings, YouTube videos) do not have available decompositions into their isolated components. They therefore cannot be used to train models using the dominant training approach.

A system that could bootstrap learning a model from audio scenes where no pre-separation into isolated sources is available would be foundational to building systems that can learn from broadly available sources of audio (e.g., audio from YouTube, or live microphone input) containing a much larger range of sounds in a much larger range of mixtures, than is possible with synthetic mixtures. This, in turn, should lead to more robust source separation models able to separate more classes of sounds in more kinds of mixtures.

The standard learning procedure for deep source separation models is in contrast to how humans learn to segregate audio scenes \cite{bregman1994auditory}: sources are rarely presented to us in isolation and almost never in ``mixture/reference'' pairs. We learn to attend to auditory scenes without ever having access to large datasets of isolated sounds. There is experimental evidence that the brain uses primitive cues (e.g., direction of arrival, repetition, proximity in pitch and time) that are independent of the characteristics of any particular sound source to perform an initial segmentation of the audio scene \cite{mcdermott2011empirical}. The brain could use such cues to separate at least some scenes to some extent, and use that information to train itself to separate more difficult scenes \cite{mcdermott2011recovering}.

In this work, we present a method to train a deep learning system %that mimics this ability and demonstrate it on the task of music/voice separation. In our system, we first separate mixtures using 
from the output of multiple primitive separation algorithms. We first combine the outputs of primitive separation algorithms in a way that out-performs any single primitive algorithm by itself. Then we use a confidence measure that is predictive of the performance of the primitive ensemble to create training data for a deep learning model. We apply our method to train the network on a large set of music recordings from YouTube \textit{without} using ground truth separation.

\begin{figure}
    \centering
    \includegraphics[width=1.0\linewidth]{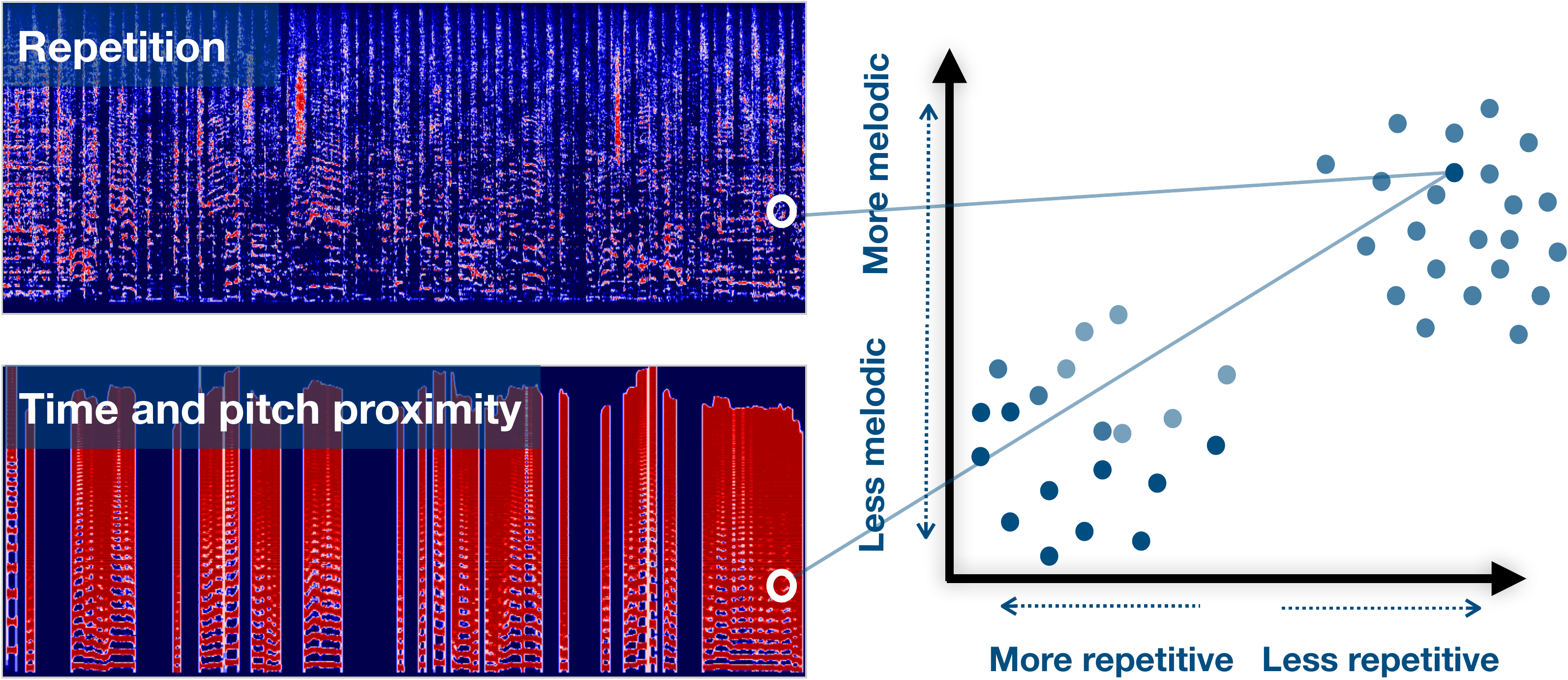}
    \caption{The process for making a primitive embedding space. A set of primitive algorithms is run on the mixture. Each algorithm produces a mask with values between 0 (blue) and 1 (red) that indicates how it is segmenting the auditory scene. Here, we show primitive clustering for two primitives. Together, they map each time-frequency point to a 2D embedding space, shown on the right. The marked point was classified by the two primitives as melodic and not repetitive, indicating that it likely belongs to the vocals estimate.}
    \label{fig:primitive_clustering}
\end{figure}

There have been a number of approaches in recent years to performing source separation using primitive cues. These include algorithms to segment the auditory scene based on common-fate (direction of moving frequency complexes) \cite{seetharaman2017music}, melodic contours \cite{salamon2012melody}, and direction of arrival \cite{fitzgerald2016projet}. Algorithms for separating repeating from non-repeating elements of the auditory scene \cite{rafii2013online,liutkus2012adaptive} and separating harmonic from non harmonic elements \cite{fitzgerald2010harmonic} have also been implemented. We build on this prior work, using similar primitive algorithms in our ensemble of primitives.

A number of ensemble approaches have also been tried. Le Roux et al.~\cite{LeRoux2013WASPAA10} proposed to use ensemble learning for speech enhancement, combining the time-frequency masks corresponding to outputs of multiple conventional enhancement algorithms using a shallow classifier. The classifier however needs to be trained in a supervised way. Another method of note is kernel additive modeling (KAM) \cite{liutkus2014kernel}, which puts primitive separation algorithms into a framework where elements of the time-frequency representation are clustered using a set of primitive-specific proximity kernels. Their approach, however, assumes each source can be separated using a single primitive and does not provide a way of estimating confidence in that primitive. Other approaches \cite{mcvicar2016learning, bach2006learning, manilow2017predicting} learn how to ensemble multiple primitives together but require ground truth to train the mechanism which combines the output of the primitives. Our system does not require ground truth at any stage of the training process.

We build on state-of-the-art deep learning source separation models \cite{luo2017deep, seetharaman2019class} that take an approach based on deep clustering
\cite{hershey2016deep}. Such models can be trained to separate different sources but require ground truth for training. Here, we eliminate the need for ground truth in the training process by learning directly from primitives. Previous efforts \cite{drude2019unsupervised, seetharaman2019bootstrapping, tzinis2019unsupervised} to train source separation models without ground truth all learn models to separate speech in stereo mixtures using only the direction of arrival primitive to generate training data. Our work provides a framework for using multiple primitives, rather than a single primitive, in a mediated ensemble to train the deep model. We apply our method to music source separation. We also propose a novel technique for augmenting the training data for the model via remixing sources discovered by the primitive ensemble.

\section{Primitive Clustering}

The first component of our system is \textit{primitive clustering}, a method for combining multiple primitive-based algorithm (Figure \ref{fig:primitive_clustering}).
Given a music mixture, each primitive-based algorithm produces a soft time-frequency mask that separates the mixture into an accompaniment estimate and a vocals estimate. This soft time-frequency mask maps each time-frequency point in the mixture to a number between $0$ (accompaniment) and $1$ (vocals). The masks produced by each primitive can be placed into a joint primitive embedding space, where each dimension of the embedding space contains the soft mask value for a time-frequency point according to one of the primitives. Given $D$ primitives, each time-frequency point $X(t,f)$ will thus be represented as a $D$ dimensional vector.

To turn the embedding space into a soft mask that takes into account all of the decisions made by each algorithm, we use an approach related to soft K-Means clustering \cite{jain2010data}. Instead of determining the means of the two clusters from the data (as in K-Means), we fix them to be the points $\mu_0 = [0]^D$ and $\mu_1 = [1]^D$. These points are where the primitive separation algorithms all strongly agree on how to assign a time-frequency point to either the accompaniment (0) or vocals (1). We then use the distance of the primitive embedding of every time-frequency point to $\mu_0$ and $\mu_1$ to calculate the soft mask for each source $M_{k}(t, f)$ at each point as follows:
\begin{equation}
\label{eq:chap2:softmax_pcl}
M_{k}(t, f) = 
    \frac
        {e^{-\beta \mathcal{D}(
            \mathcal{F}(X(t, f)),
            \mu_k)
        }}
        {\sum_j e^{-\beta \mathcal{D}(
            \mathcal{F}(X(t, f)),
            \mu_j)
        }
    }
\end{equation}
where $\mathcal{D}(x, y)$ is the Euclidean distance between points $x$ and $y$ and $\mathcal{F}$ is a function that maps a time-frequency point into the primitive embedding space. This maps distances in the embedding space to values between $0$ and $1$. $\beta$ is a hyperparameter that controls the hardness of the decisions made by the clustering algorithm. In this work we use $\beta=5.0$. 

\section{Bootstrapping via remixing}

Our goal is to train a deep learning model directly from mixtures without ground truth by generating training data via an ensemble of primitive separation algorithms. One simple way to do this would be to apply the primitives to a large set of mixtures and train the network directly from the time-frequency labels they generate for each mixture, as was done in prior work applied to the output of a single primitive \cite{seetharaman2019bootstrapping,tzinis2019unsupervised,drude2019unsupervised}. If, however, the primitive separations fail on some of these mixtures then the network will be trained with noisy data. To mitigate this, we use two strategies: a confidence measure that can be used to eliminate failure cases in the training process, and a new data augmentation process to generate additional higher-quality training data.

\subsection{Confidence measure}

\begin{figure}
    \centering
    \includegraphics[width=1.0\linewidth]{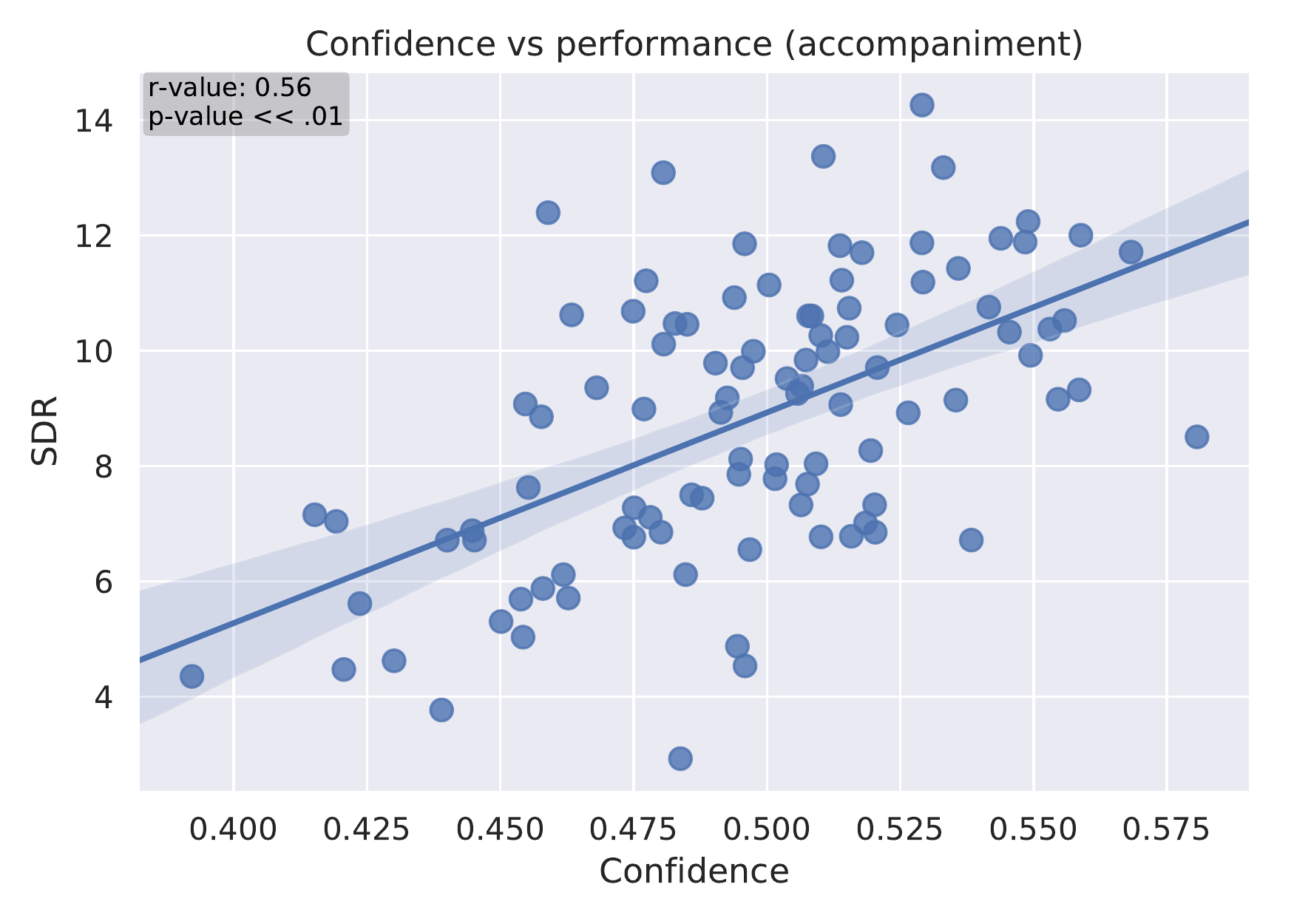}
    \caption{Relationship between confidence measure and actual performance for music mixtures, using primitive clustering to separate each mixture. Each of the $100$ points represents the accompaniment (non vocals) produced by primitive clustering on one of the 100 MUSDB mixtures. The blue line is the line of best fit found via linear regression. The p-value and r-value for the regression are overlaid.}
    \label{fig:primitive_clustering_conf_rel}
\end{figure}

The goal of the confidence measure is to estimate the performance of a source separation algorithm without access to ground truth. In every clustering-based algorithm, time-frequency points are mapped to an embedding space in which the clustering is performed. The core insight behind the confidence measure is that the structure of this embedding space is related to the performance of the algorithm. By analyzing the embedding space, we can estimate the performance of any clustering-based source separation algorithm without the need to compare to ground truth sources. The confidence measure has two parts: the silhouette score and posterior strength. Posterior strength and silhouette score are combined by multiplying them together so that if either is low, the entire confidence measure is low. 

\subsubsection{Silhouette score}
The silhouette score \cite{rousseeuw1987silhouettes} produces a score for every point in a dataset that corresponds to how well that point is clustered. To compute the silhouette score, let us first assume we have a partition of dataset $X = \bigcup_{k=1}^K C_k$ into $K$ clusters. For a data point $x_i$ in cluster $C_k$, we compute the following terms:
\begin{align*}
a(x_i) &= \frac{1}{|C_k| - 1} \sum_{x_j \in C_k, i \neq j} d(x_i, x_j), \\
b(x_i) &= \min_{o \neq k} \frac{1}{|C_o|} \sum_{x_j \in C_o} d(x_i, x_j).
\end{align*}
$a(x)$ is the mean distance (using a distance function $d$) between $x_i$ and all other points in $C_k$, and $b(x)$ is the mean distance between $x_i$ and all the points in the nearest cluster. The silhouette score is defined as
\begin{equation}
    \label{eq:chap3:silhouette}
	s(x) = \frac{b(x_i) - a(x_i)}{\max (a(x_i), b(x_i))} \, \text{ if } |C_k| > 1,
\end{equation}
and $s(x_i) = 0$ if $|C_k| = 1$. $s(x)$ ranges from $-1$ to $1$.

To apply the silhouette score to primitive clustering we do not use the soft assignment for each time-frequency point from Eq. \ref{eq:chap2:softmax_pcl}. Instead, we assign each point to the cluster (vocals or accompaniment) whose center it is closest to. 

Computing the silhouette score for every point in a typical auditory scene (often millions of points) is intractable. Instead, we sample $N$ points from the embedding space and compute the silhouette score of each point in the sampled subset. With a relatively small $N$, we can approximate the mean silhouette score efficiently. We set $N=1000$, and sample these $N$ points from the loudest $1$\% of time-frequency bins in the mixture spectrogram, to focus our estimate on elements that are perceptually prominent. 

\subsubsection{Posterior strength}

For every point $x_i$ in a dataset $X$, the clustering algorithm produces soft assignments $\gamma_{ik} \in [0, 1]$ that indicates the membership of the point $x_i$ in some cluster $C_k$. $\gamma_{ik}$ is also called the \textit{posterior} of the point $x_i$ in regards to cluster $C_k$. The closer $\gamma_{ik}$ is to $0$ (not in the cluster) or $1$ (in the cluster), the more confident the assignment of that point. For a point $x_i$ with corresponding $\gamma_{ik}$ for $k \in [0, 1, ..., K]$, we compute:
\begin{equation}
    \label{eq:chap3:posterior_confidence}
    P(x_i) = \frac{K (\max_{k \in [0, ..., K]} \gamma_{ik}) - 1}{K - 1}
\end{equation}
where $K$ is the number of clusters, and $P(x_i)$ is the \textit{posterior strength}, as it captures how strongly a point is assigned to any of the $K$ clusters. This equation maps points that have a maximum posterior of $\frac{1}{K}$ (equal assignment to all clusters) to confidence $0$, and points that have a maximum posterior of $1$ to confidence $1$. To compute a single score for this measure, we take the mean posterior strength across the top 1\% of points by loudness.

\begin{figure}[t]
    \centering
    \includegraphics[width=1.0\linewidth]{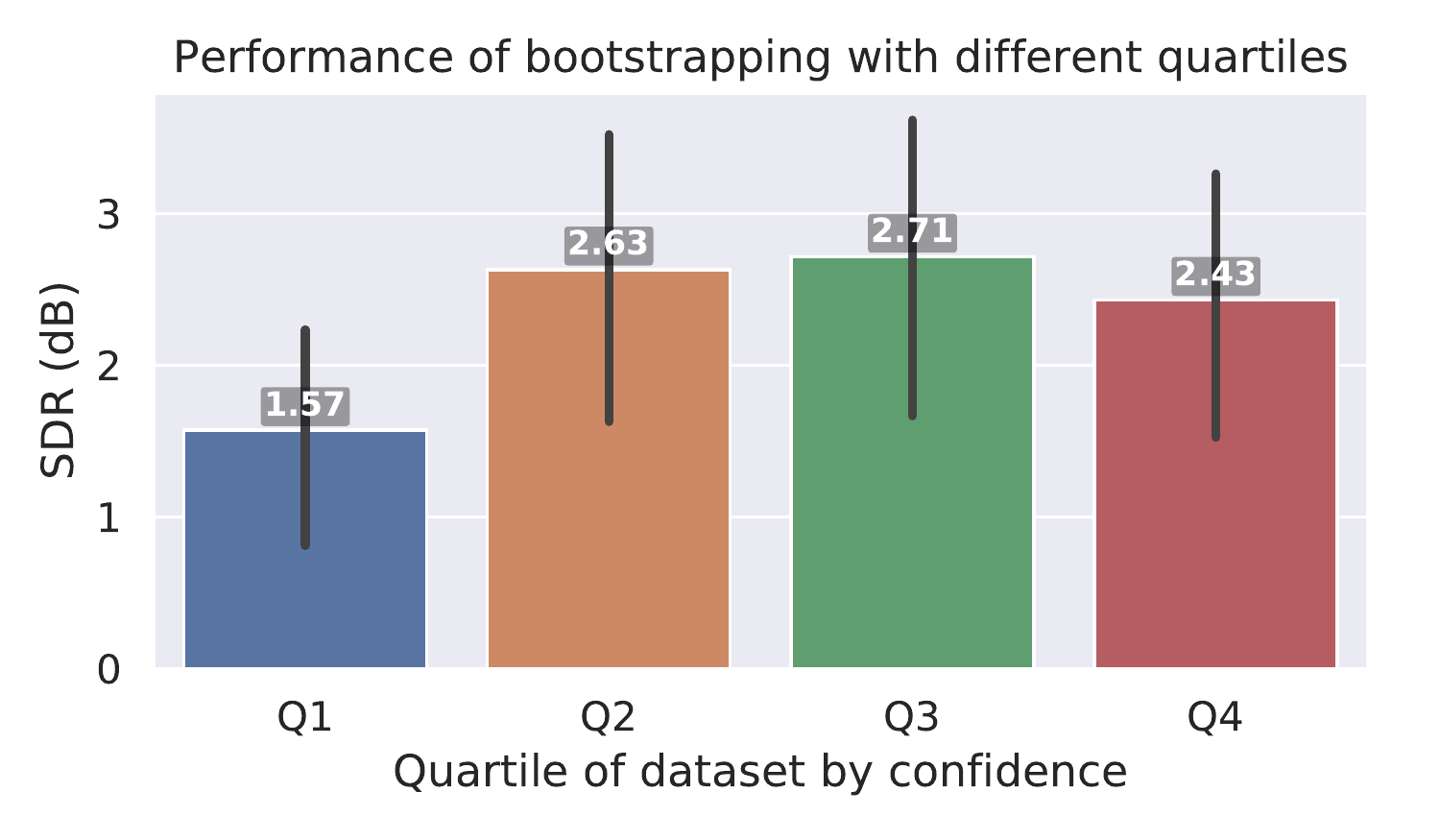}
    \caption{Performance of bootstrapping when the training mixtures are constructed with different quartiles of the training set, organized by the confidence measure. Q1 indicates that the network was bootstrapped from separated sources in the lowest quartile by confidence. Q2 indicates the next quartile, and so on. The network bootstrapped from the lowest quartile has poor performance, indicating that the confidence measure is good at ruling out low-quality sources.}
    \label{fig:effect_of_conf}
\end{figure}

\subsection{Putting it all together}

We first apply primitive clustering to a large set of music mixtures. For each mixture that we separate, we have an associated confidence measure that indicates how well we think the primitives performed on that mixture. This generates a large set of separated source estimates, each with an associated confidence in the quality of the estimate. We remove low-confidence source estimates from this set, leaving a smaller set of high-confidence estimates. High-confidence source estimates from different mixtures are then combined to create new mixtures (i.e. vocal estimate from mixture A and accompaniment from mixture B) used to train the deep network. 

This remixing procedure augments the amount of data that is used to train the deep network. If we were to learn only from the original mixtures, this greatly limits the amount and diversity of data available to learn from. Additionally, the separated sources are separated via primitives, which were successful due to characteristics of the original mixture. Remixes using source estimates from two different mixtures often do not share the mixing characteristics of the mixes either source estimate came from. This forces the deep network to learn a different way to separate the two sources than was used by the primitives applied to the original mixtures.

\section{Experiments}
We aim to answer the following questions in our experiments:

\begin{enumerate}[itemsep=0cm]
    \item Is the confidence measure predictive of the performance of primitive clustering?
    \item How using the confidence measure to select training examples impact the effectiveness of the bootstrapped separation model?
    \item How does a bootstrapped model perform relative to a model trained using ground truth isolated sources, and relative to the primitive separation algorithms used to train the model?
\end{enumerate}

\begin{figure}[t]
    \centering
    \includegraphics[width=1.0\linewidth]{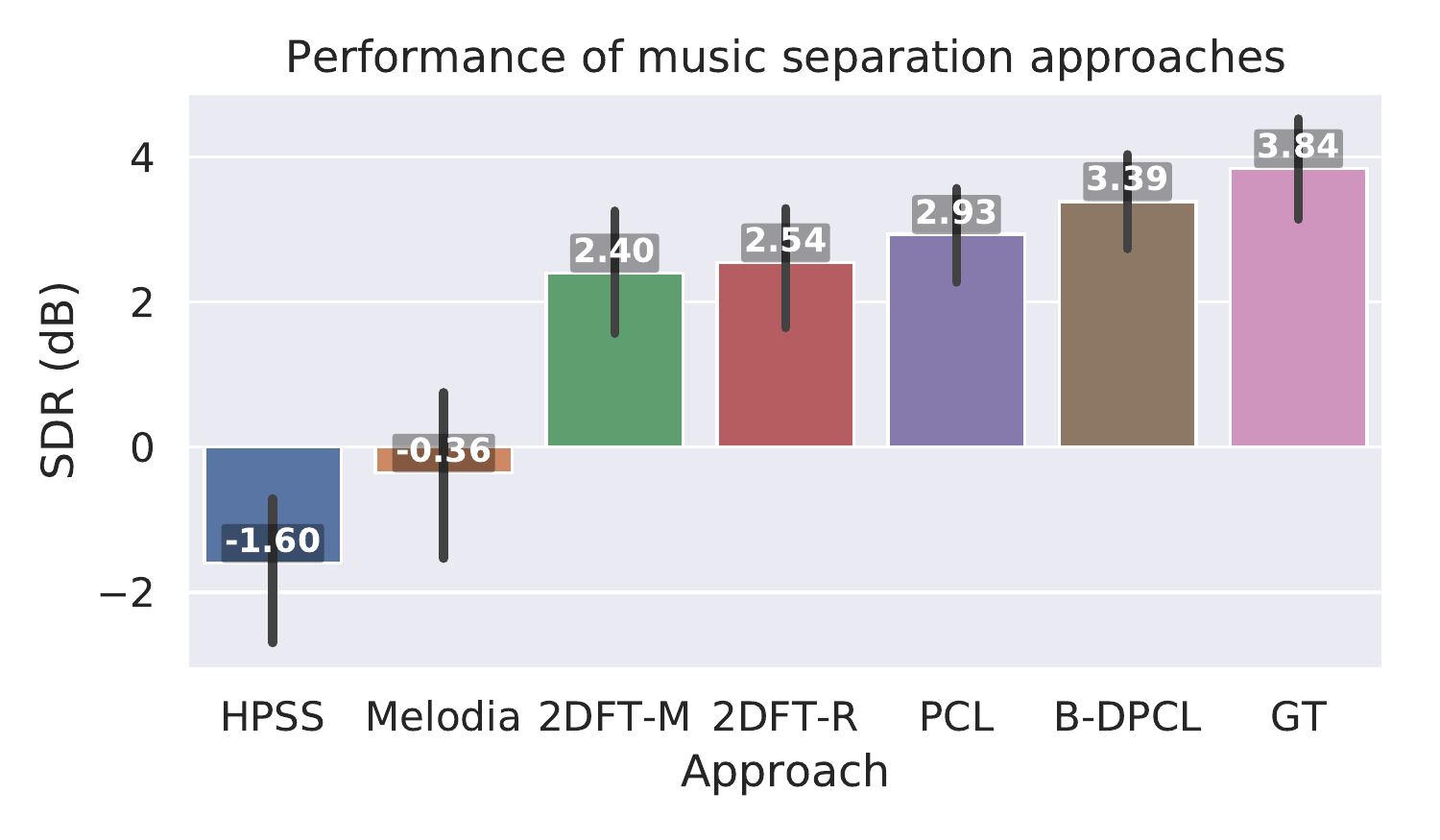}
    \caption{Separation results on test data. The bootstrapped model (B-DPCL) significantly outperforms primitive clustering (PCL), which was used to train it, but falls short of the performance of a model trained on ground truth separation (GT). Higher values are better.}
    \label{fig:all_approaches}
\end{figure}

\subsection{Data}
To create training mixtures for the ground truth network, we remixed the stems of the $100$ training tracks from the MUSDB dataset \cite{musdb18}  to create $20000$ new mixtures for training and $5000$ mixtures for validation. This was done by taking random $15$ second snippets from some accompaniment stems (the sum of the bass, drums, and other stems) and some vocal stem in the dataset and remixing them at a random signal-to-noise ratio between $-2.5$ and $2.5$ dB. Each mixture had a sample rate of $44100$ Hz, and was transformed using an STFT with window length $2048$ and hop $512$.

In our primitive clustering ensemble, we used the following primitives: micro-modulation (2DFT-M) (2DFT-R) \cite{seetharaman2017music}, repetition \cite{seetharaman2017music}, time/pitch proximity (Melodia) \cite{salamon2012melody}, and timbre as embodied in harmonic/percussive timbres (HPSS) \cite{fitzgerald2010harmonic}. We created training data to bootstrap a network from segmentations produced by primitive clustering as follows. Given a set of musical mixtures, we first split each mixture up into $30$ second segments, with $15$ second overlap between segments, filtering out quiet segments. We applied primitive clustering to the remaining segments, resulting in a set of separated vocals and accompaniment sources. These separations were remixed to create $20000$ training mixtures. This procedure was applied to two sets of mixtures. The first was the same the $100$ mixtures from MUSDB that were used to train the ground truth network. The second set adds $831$ songs across $4$ genres (oldies, opera, pop, and rock) downloaded from YouTube. Both sets contain $20000$ remixes, but the remixes from the set including YouTube audio are more diverse.

\subsection{Training a network}

Each deep network trained was a deep clustering network \cite{hershey2016deep}, consisting of a stack of $4$ BLSTM layers that output 20-dimensional embeddings with sigmoid activation and unit-normalization. Each network was trained using sequences of $400$ frames. Each network was trained for $50$ epochs with the ADAM \cite{kingma2014adam} optimizer with a learning rate of 2e-4. Dropout of $0.3$ was applied to each BLSTM during training. We used the deep clustering objective with magnitude weights \cite{wang2018alternative}. We evaluated each separation algorithm on the MUSDB test set ($50$ tracks) using scale-dependent source-to-distortion ratio (SDR) \cite{LeRoux2018SISDR}.

\section{Results}
Figure \ref{fig:primitive_clustering_conf_rel} shows the relationship between the confidence measure and SDR on the MUSDB training set for accompaniment sources. A linear fit  between the two has an r-value of $0.56$. This shows that the confidence measure is broadly predictive of performance for primitive clustering on a set of musical mixtures. 

We now study the effect using the confidence measure to filter the training data produced by primitive clustering has on the performance of the learned model. Each mixture in the MUSDB train set was separated into vocals and accompaniment by primitive clustering. Each separations has an associated confidence (see Figure \ref{fig:primitive_clustering_conf_rel}). We split these separations into four quartiles, from Q1 (source estimates with the lowest confidence score) to Q4 (estimates with the highest confidnece). We applied the same remixing procedure to the sources in each quartile, creating 20000 training mixtures used to train four network - one per quartile. Figure \ref{fig:effect_of_conf} shows the performance for these four networks. The network  bootstrapped from the lowest quartile has poor performance. Those trained on the other quartiles have better performance, indicating that the confidence measure for primitive clustering (which does not depend on ground truth) can be used to filter out low-quality training examples.

Figure \ref{fig:all_approaches} shows the results of each separation algorithm on this test set. Primitive clustering (PCL) outperforms the primitives that went into it: harmonic/percussive timbre (HPSS), time and pitch proximity (Melodia), micro-modulation (2DFT-M), and repetition (2DFT-R). Following the results in Fig.~\ref{fig:effect_of_conf}, we trained a bootstrapped network from the larger set that included the YouTube data, excluding the bottom quartile of separated sources by confidence. This network achieves $3.39$ dB SDR, out-performing its teacher (PCL) and nearing the performance of the ground truth network.

The amount of data that is used to create the mixtures has a considerable effect on the performance of the bootstrapped networks. The model that is trained with only mixtures created from the MUSDB training set has SDR lower than the primitive method ($2.71$ vs $2.93$). Increasing the amount of data using the YouTube mixtures increases the performance by $0.68$ dB SDR. The performance of the bootstrapped networks still falls short of the ground truth network. This gap in performance could perhaps be addressed by adding more data. However, it is much easier to add training data for the bootstrapped network than for one trained on traditional ground truth separated sources, since any audio mixture can be used as training data.

\section{Conclusion}
We have presented a method for learning to separate sounds directly from mixtures without ground truth by using primitive auditory grouping principles. To do so, we developed a method for combining multiple primitive algorithms called primitive clustering. We use a simple method for estimating (without need for ground truth) the performance of not just primitive clustering, but any clustering-based separation algorithm. This lets us apply primitive clustering to a large set of mixtures, sift through the separated sources to find ones that are likely to be of high quality, and then remix those separated sources into training data that can be used to bootstrap a deep network from audio data where we do not have ground truth isolated sources. This opens the door to deep separation models that can continuously learn in the wild via bootstrapping.

% -------------------------------------------------------------------------
% Either list references using the bibliography style file IEEEtran.bst
\bibliographystyle{IEEEtran}
\bibliography{refs.bib}

\end{sloppy}
\end{document}